\documentclass{article}

\usepackage{makeidx}  % allows for indexgeneration
\usepackage{graphicx}
\usepackage{subfigure}
\usepackage{lineno}

\usepackage[T1]{fontenc}
\usepackage[english]{babel}

\usepackage{amssymb}
\usepackage{amsmath}
\usepackage{amsfonts}

\usepackage{bbm}

\title{Some open questions on morphological operators and representations in the deep learning era \\
	\emph{A personal vision.}~\footnote{Notes for Keynote at DGMM'2021 (IAPR International Conference on Discrete Geometry and Mathematical Morphology), Uppsala University, Sweden, May 24-27, 2020.}}

\author{
	Jes\'us Angulo \\
	MINES  ParisTech,  PSL-Research University\\
	CMM-Centre de Morphologie Math\'{e}matique, France \\
	\texttt{jesus.angulo@mines-paristech.fr} \\
}

\date{March 2021}

\begin{document}

\maketitle              % typeset the title of the contribution

\begin{abstract}
\emph{“Work on deep learning or perish”}: folklore wisdom in 2021. 

\bigskip

During recent years, the renaissance of neural networks as the major machine learning paradigm and more specifically, the confirmation that deep learning techniques provide state-of-the-art results for most of computer vision tasks has been shaking up traditional research in image processing.  The same can be said for research in communities working on applied harmonic analysis, information geometry, variational methods, etc. For many researchers, this is viewed as an existential threat. On the one hand, research funding agencies privilege mainstream approaches especially when these are unquestionably suitable for solving real problems and for making progress on artificial intelligence. On the other hand, successful publishing of research in our communities is becoming almost exclusively based on a quantitative improvement of the accuracy of any benchmark task. 

\bigskip

As most of my colleagues sharing this research field, I am confronted with the dilemma of continuing to invest my time and intellectual effort on mathematical morphology as my driving force for research, or simply focussing on how to use deep learning and contributing to it. The solution is not obvious to any of us since our research is not fundamental, it is just oriented to solve challenging problems, which can be more or less theoretical. Certainly, it would be foolish for anyone to claim that deep learning is insignificant or to think that one's favourite image processing domain is productive enough to ignore the state-of-the-art. I fully understand that the labs and leading people in image processing communities have been shifting their research to almost exclusively focus on deep learning techniques. My own position is different: I do think there is room for progress on mathematically grounded image processing branches, under the condition that these are rethought in a broader sense from the deep learning paradigm. Indeed, I firmly believe that the convergence between mathematical morphology and the computation methods which gravitate around deep learning (fully connected networks, convolutional neural networks, residual neural networks, recurrent neural networks, etc.) is worthwhile.  

\bigskip

The goal of this talk is to discuss my personal vision regarding these potential interactions. Without any pretension of being exhaustive, I want to address it with a series of open questions, covering a wide range of specificities of morphological operators and representations, which could be tackled and revisited under the paradigm of deep learning. An expected benefit of such convergence between morphology and deep learning is a cross-fertilization of concepts and techniques between both fields. In addition, I think the future answer to some of these questions can provide some insight on understanding, interpreting and simplifying deep learning networks.

%\keywords{}
\end{abstract}
%======================================================================

% ----------------------------------------------------------------
\section{Mathematical morphology is powerful and still attractive despite its age}

Mathematical morphology is \emph{not only} a mathematical theory of shape. Its corpus nowadays provides a vast theoretical and practical machinery to address fundamental problems arising from the fields of computer vision and structured-data analysis.

Unfortunately, the significant scope of morphology is overshadowed by its widely use as mainly a post-processing tool to regularize binary images as well as the progressive shift of interest of practitioners towards deep learning techniques, which require little knowledge on image processing theory and provide impressive results. A ``win-win'' game. In the case of theorists, a progressive fading of interest has been caused by at least three possible reasons: i) the apparent exotic mathematical formulation of morphology, ii) a theoretical apparatus which is aged, or worse,  already depleted it of new discoveries, iii) the perception of morphology as a useless theory for the future of signal/image and data processing since, once again, deep learning can solve everything  and the mathematics underlying deep learning cannot interact with the algebraic and geometrical formulation of morphology. I challenge these three arguments. I believe, on the contrary, that the arrival of real progress on artificial intelligence based on deep learning provides a mind frame to push the boundaries of morphology and to prove that it is one of the appropriate nonlinear machineries to address some open issues on understanding, interpreting and simplifying deep learning networks. But also to introduce new layers and architectures inspired from morphological operators and representations. Let me enumerate the major themes on mathematical morphology which are relevant in the context of this talk:
\begin{itemize}
	\item An abstract algebraic formulation of the theory on complete lattices, which requires very little assumptions to be instantiated into a specific lattice structure of the space of interest.
	\item A common representation theory for the Boolean and the semicontinuous function cases, in which, for instance, any translation-invariant increasing, upper semicontinuous operator can be represented exactly as a minimal superposition of morphological erosions or dilations.
	\item An intimate relationship with the random set theory via the notion of Choquet capacity from stochastic geometry.
	\item Strong connections with idempotent mathematics (max-plus and max-min algebra and calculus) and tropical geometry.
	\item Continuous models which correspond to Hamilton--Jacobi PDEs, relevant also in optics and optimal control.
	\item A powerful extension to the case of morphology on groups, which bring a proper dealing with space symmetries and provide equivariant operators to the groups of transforms relevant in computer vision.
	\item Multiscale operators and semigroups formulated in Riemannian, metric and ultrametric spaces.
	\item Multiple morphological representations that provide a rich family of shape-based and geometrical descriptions and decompositions: skeletons, pattern spectra and size distributions, topological description of functions using maxima-minima extinction values, etc.
	\item A privileged mathematical tool for Lipschitz characterization and regularization.
	\item A counterpart of the perceptron which yields to the scope of morphological neural networks, morphological associative memories.
\end{itemize}

The previous list is not exhaustive and of course, it is  based on my personal research interests. Nevertheless, I do believe it illustrates which I mean by a central position on the field of nonlinear mathematics for visual computing.

\section{A selection of themes where morphology and deep learning can meet}

I discuss now in a rather informal way a few fields of potential interaction between deep learning and mathematical morphology and the questions which arise from. The bibliography is not exhaustive since I am covering a large scope of topics. Some of the subjects that I mention below are already the object of current research. Due to prospective nature of these reflections, this should be considered more like a personal roadmap than a systematic review of the state-of-the-art.

%%%%%%%%%%%%%%%%%%%%%%%%%%%%%%%%%%%%%%%%%%%%%%%%%%%%%%%%%%%%%%%%%
\subsection{Lattice theory and algebraic models for deep learning}

Current mathematical models for deep learning networks are based on approximation theory and harmonic analysis~\cite{Wiatowski18,Daubechies19}. Other approaches explore the relevance of tropical geometry~\cite{Maragos19} to describe networks with Rectified Linear Units (ReLUs)~\cite{Zhang18,Arora18}. The Matheron-Maragos-Banon-Barrera (MMBB)~\cite{Matheron75,Maragos89,Banon91} representation theorems provide an astonishing general formulation for any nonlinear operator between complete lattices, based on combinations of infimum of dilations and supremum of erosions. The theory is relevant when the basis (minimal kernel) of the operators can be learnt. In the case of non-increasing or non-translation-invariant operators the constructive decomposition of operators become more complex but still it would based on basic morphological dilation, erosion, anti-dilation and anti-erosion which can be the minimal bricks to construct architectures of networks which mimic the MMBB representations.

\begin{quote}
	How effective MMBB networks would be to learn the minimal basis of structuring functions approximating any nonlinear image transform? How the idea of hierarchical architectures from deep learning can be used in the case of MMBB networks?
\end{quote}  

\begin{quote}
	Can MMBB networks be combined with standard layers in deep learning pipelines, providing relevant learnable models?
\end{quote}  

Any network architecture combining convolution, down/up-sampling, ReLUs, etc. could be seen at first sight as incompatible with lattice theory formulation. In fact, as it was shown by Keshet~\cite{Keshet02,Keshet03}, low-pass filters, decimation/interpolation, Gaussian/Laplacian pyramids and other typical image processing operators, admit an interpretation as erosions and adjunctions in the framework of (semi)-lattices. In addition, max-pooling and ReLUs are just dilation operators. The notion of deepness or recurrence in a network can be seen as the iteration of basic operators, which yields to questions on the convergence to idempotency or, at least, to study order stability in the corresponding lattice~\cite{Hejmans92}.

\begin{quote}
	What kind of unified algebraic models, integrating standard layers and morphological layers can be used to mathematically study deep learning architectures? Is there any information on order continuity, on invariance and fixed points, on decomposition and simplification, etc., which can be inferred from these unified algebraic models?  
\end{quote}  

\begin{quote}
	What is the expressiveness of deep MMBB networks and the hybrid deep networks?
\end{quote}  

This last question is related to study of the capacity of neural networks to be universal approximators for smooth functions. For instance, both maxout networks~\cite{Goodfellow13} and max-plus networks~\cite{Zhang19} can approximate arbitrarily well any continuous function on a compact domain. The proofs are based on the fact that~\cite{Wang04} continuous piecewise linear (PWL) functions can be expressed as a difference of two convex PWL functions, and each convex PWL can be seen as maximum of affine terms. Alternative theory by Ovchinnikov~\cite{Ovchinnikov01,Ovchinnikov02} shows that a PWL (or a smooth) function can be represented as a max-min polynomial of its linear components, and the theory is also valid in a Boolean representation. The representation formulas by Ovchinnikov are equivalent to MMBB theorems which justify the potential interest of the latter. Tropical formulation of ReLU networks has shown that a deeper network is exponentially more expressive than a shallow network~\cite{Zhang18}. To explore the expressiveness of complex morphological networks with respect to the deepness is therefore a fundamental relevant topic.

%%%%%%%%%%%%%%%%%%%%%%%%%%%%%%%%%%%%%%%%%%%%%%%%%%%%%%%%%%%%%%%%%
\subsection{Lipschitz regularity in neural networks}

Lipschitz regularity has been proven to be a fundamental property to deal with robustness of the predictions made by deep neural networks when their input is subject to an adversarial perturbation~\cite{Goodfellow15}. This mathematical topic of Lipschitz regularity is quite important in deep learning since adversarial attacks against machine learning models are a proof of their limited resilience to small perturbations. Training neural networks under a strict Lipschitz constraint is useful also for generalization bounds and interpretable gradients~\cite{Tsipras18}. By the composition property of Lipschitz functions, it suffices to ensure that each individual affine transformation or nonlinear activation is 1-Lipschitz. That can be achieved by constraining the spectral norm of the weights in the layers: for instance, maintaining during the training weight matrices of linear and convolutional layers to be approximately Parseval tight frames (extensions of orthogonal matrices to non-square matrices)~\cite{Cisse17}. 
This approach satisfies the Lipschitz constraint, but comes at a cost in expressive power~\cite{Huster18}. 
Other techniques replace the ReLU layers by more elaborated functions based of ordering the inputs and computing max-min operations (GroupSort activation function)~\cite{Anil19}.
Morphological operators using multiscale convex structuring functions are a powerful tool to deal with Lipschitz extension of functions~\cite{Angulo14}, which is connected to Lasry-Lions regularization~\cite{Lasry86}.

\begin{quote}
What kind of Lipschitz morphological layers can be introduced into deep networks to control their Lipschitz constant and therefore their regularity? Can these morphological regularizing layers replace the standard nonlinearities like pooling+ReLU without degrading their expressiveness?
\end{quote}

The later question is also the object of related work~\cite{Cohen19a}, where it has been proven that Lipschitz constraint models using FullSort activation functions are universal Lipschitz function approximators.

%%%%%%%%%%%%%%%%%%%%%%%%%%%%%%%%%%%%%%%%%%%%%%%%%%%%%%%%%%%%%%%%%
\subsection{Group equivariance and integration of data symmetries and topology}

Motivated by the Gestalt pattern theory, some studies investigated by synthetic experiments the ability of deep learning to infer simple (at least for human) visual concepts, such as shape or symmetry, from examples~\cite{Yan2017}. Humans can often infer a semantic geometric/morphological concept quickly after looking at only a very small number of examples; on the contrary, deep convolutional neural networks approximate some concepts statistically, but only after seeing many (thousands or millions) more examples. It seems reasonable that the use of morphological layers which deal more naturally with the notion of shape could improve some visual taks.

\begin{quote}
Are networks integrating morphological layers more ``intelligent'' (i.e., requires less training samples) than standard deep learning architectures to learn tasks inspired from Gestalt pattern theory?
\end{quote}

I think in particular about the potential role to be played by the theory of group morphology~\cite{Roerdink00}, which extends the construction of morphological operators to be invariant under, for instance, the motion group, the roto-translation group, the affine group, the projective group, etc.  Indeed, the notion of group equivariance of networks~\cite{Cohen16} is central nowadays in the field of deep learning. It provides a sound approach to deal with the explicit introduction of the desired symmetries into the network, without the need to approach them by means of costly techniques such as data augmentation.

A combinatorial shape problem called generalized Tailor problem~\cite{Roerdink96}, which is connected to the one of finding the decomposition of any shape according to a set of templates, can be relevant to assess the interest of group morphology in deep learning based part-based decomposition. 

\begin{quote}
How efficiently can  the generalized tailor problem be solved using networks inspired from the morphological iterative algorithm?
\end{quote}

Another interesting problem is to design models for machine learning tasks defined on sets~\cite{Zaheer18}. In contrast to traditional approaches, which operate on fixed dimensional vectors, the idea is to consider objective functions on sets that are invariant to permutations. The issue is also relevant on graphs. This equivariance to the permutation of the elements of the input requires specific pooling strategies across set-members. In mathematical morphology and discrete geometry, there are compact representations of sets by a minimal number of points, typically based on the notion of skeleton, maxima/minima of the signed distance function, etc. Other shape morphological representations such as shape-size distribution from granulometries, provided set descriptors which are invariant to many transforms of the sets. It was proved in the past the interest of skeletons and shape-size representations when they are combined with neural networks~\cite{Yang93}.

\begin{quote}
Are morphological compact representations of sets more efficient to deal with permutation invariance in machine learning? Can we introduce loss functions based on morphological descriptions which are inherently permutation-invariant and scale-invariant?
\end{quote}

In a similar way, considering that an image is a function whose relevant information is associated to its topology, namely the location of the maxima/minima and the features associated to them, provides a representation which is invariant to many isometries. Dealing with maxima and minima can be addressed using morphological representations based on residues of morphological reconstruction (and iterative algorithm) and the appropriate markers~\cite{Vincent93}. The idea of topology-preserving has been considered for the problem of deep image segmentation~\cite{Hu19}, basically to learn continuous-valued loss functions that enforces a segmentation to have the same topology as the ground truth. That is done using the notion of persistence diagrams from computational topology~\cite{CohenSteiner07}. However, the integration of computational topology and deep learning is not natural. 

\begin{quote}
Can we learn persistence (extinction values) features to be used in topological image classification and segmentation using morphological operators based on reconstruction? Which architecture to be used for the iterative-based reconstruction (residual, recurrent, other)? 
\end{quote}

%%%%%%%%%%%%%%%%%%%%%%%%%%%%%%%%%%%%%%%%%%%%%%%%%%%%%%%%%%%%%%%%%
\subsection{Interpretability and ``small'' parametric models}

Due to the black-box nature of deep learning, it is inherently difficult to understand which aspects of the input data are contributing to the decisions on a complex network. It is also difficult to identify which combinations of data features are appropriate in the context of the deployment of networks as a decision support system in critical domains. Understating better by humans why a deep neural network is taking a particular decision is the object of the so-called explainable deep learning~\cite{Xie20}. 

A way to move towards explainable networks is to have layers which are easy to interpret. For instance, if a part of the network is learning patterns, as in a template matching problem, and those patterns are easy to visualize, this part of the networks can be explainable. One of the most studied and rather simple morphological operator (i.e., the intersection of an erosion and an anti-dilation), the hit-or-miss transform, can be seen as a powerful template matching approach. 

The use of the hit-or-miss transform as part of a neural network for object recognition was pioneered in~\cite{Won97} and some recent attempts of extending its use in the context of deep neural networks are promising~\cite{Islam20}. However, to have robust to noise template detection~\cite{Bloomberg00} or the multiple ways to extend the hit-or-miss transform to grey-scale images~\cite{Khosravi96,Naegel07}, as well as the fact that the patterns to be matched can appear at different scales or at different rotations, yield interesting topics to be explored. 

\begin{quote}
What is the best formulation for the hit-or-miss transform to provide easy learnable and robust template extraction layers? How efficient is the integration of hit-or-miss layers into a complex architecture of deep learning? Only as the first layers? 
\end{quote}

\begin{quote}
How to deal with equivariance in pattern detection by means of group morphology-based hit-or-miss transforms? 
\end{quote}

\begin{quote}
What other morphological template extraction operators~\cite{Ronse96} are relevant as interpretable layers in deep learning?
\end{quote}

An alternative in the quest for a better interpretability of deep learning is to replace regular convolutional neural networks filters by parametric families of canonical or well-known filters and scales spaces. That reduces significantly the number of parameters and make them more interpretable. These hybrid approaches are constructed by coupling parameterized scale-space operations in cascade~\cite{Lindeberg21}, or circular harmonics banks of filters~\cite{Worrall17} or Gabor filters~\cite{Luan18}, etc. with other neural networks layers.

In the case of morphological operators, we can also consider the use of parametric models. For instance, by defining an architecture mimicking the notion of granulometry: a series of multiscale openings followed by a global integral pooling operator, such that the parameters to be learn would be the structuring function, which would shared by all the openings, and the scale parameter for each opening. We can also consider the interest of learning parametric structuring functions (typically quadratic ones) or to consider pipelines of multiscale dilations/erosions used to predict quantitative parameters like the fractal dimension or the Hölder exponent~\cite{Angulo21}.

\begin{quote}
Are there architectures based on parametric families of morphological multiscale operators which can be efficiently learned and provide a better interpretation on tasks of shape or texture recognition?  
\end{quote}

\begin{quote}
What parametric families of structuring functions are the most fruitful in deep learning: quadratic ones defined by a shape covariance matrix? Convex ones defined by Minkowski addition of oriented segments?
\end{quote}

%%%%%%%%%%%%%%%%%%%%%%%%%%%%%%%%%%%%%%%%%%%%%%%%%%%%%%%%%%%%%%%%%
%\subsection{Disjoint parts and disentangled representations}

%%%%%%%%%%%%%%%%%%%%%%%%%%%%%%%%%%%%%%%%%%%%%%%%%%%%%%%%%%%%%%%%%
\subsection{Image generation and simulation of microstructures}
Morphological operators are the fundamental computational tool for the characterization and simulation of random sets (for instance, the binary images associated to a random microstructure) in the theory developed by Matheron~\cite{Matheron75}. The notion of Choquet capacity of a random set relies on computing how the integral of the set changes when it is dilated or eroded by a particular structuring element. By considering specific families of structuring elements (i.e., pairs or triplets of points, segments, disks, etc.), the random set is characterized and, by working on well-studied stochastic models of random sets, the corresponding parameters of the model are fit thanks to the morphological measurements. Then, it is possible to simulate new images following the model and test if the new images have the prescribed morphological measurements. This theory has been of significant success in the characterisation and simulation of microstructure images in material sciences~\cite{Jeulin18}. 

In the field of deep learning, Generative Adversarial Networks (GANs) are an approach to generate images from an illustrative dataset~\cite{Goodfellow14,Salimans16}. GANs involve automatically discovering and learning the regularities or patterns in the input data, then the model is used to generate new examples that plausibly could have been drawn from the initial dataset. GANs consider the problem as a supervised learning framework with two sub-models: the generator model that one trains to generate new examples, and the discriminator model that tries to classify examples as either real (from the domain) or fake (generated). The two models are trained together until the discriminator model is fooled about half the time, meaning the generator model is producing plausible fake examples. The generator is typically a deconvolutional neural network, and the discriminator is a convolutional neural network.

GANs are nowadays used in many domains, where image simulation or image synthesis is required, with impressive visual results. GANs are also being explored in the field of virtual material simulation in physics~\cite{Yang18,Chun20}. The additional dimension in material science is the fact that the generated image should satisfy some physical or mathematical constraints. For instance, the model can explicitly enforces known physical invariances by replacing the traditional discriminator in a GAN with an invariance checker~\cite{Singh18}.

\begin{quote}
Can the notion of Choquet capacity play a role in the GAN discriminators to explicitly impose morphological constraints learned from the empirical data?
Can one incorporate a ``deconvolution’’ image simulation closer to the morphological random set models into the GAN generators, thanks to the use of morphological layers? 
\end{quote}

%%%%%%%%%%%%%%%%%%%%%%%%%%%%%%%%%%%%%%%%%%%%%%%%%%%%%%%%%%%%%%%%%
\subsection{Ultrametric convolutional neural networks}

Many scientific fields work on data with an underlying mathematical structure modelled as a non-Euclidean space. Some examples include social networks, sensor networks, biological networks (functional networks in brain imaging or regulatory networks in genetics), and meshed surfaces in computer vision.  Geometric deep learning is a generic term for techniques attempting to generalize structured deep neural models to non-Euclidean domains such as graphs and manifolds~\cite{Bronstein17}. The area of deep learning on graphs is particularly active~\cite{Zhang20a,Bacciu20}, with many alternative approaches seeking to generalize to graphs fundamental deep learinng notions as convolution, pooling, coding/decoding, loss functions, etc. Research topics for instance deal with the role played by the nodes/edges, the use of graph spectral techniques or kernel methods, etc.

Another discrete powerful setting for structured data (or unstructured data which can be first embedded into a graph) are the hierarchical representations associated to dendrograms (rooted trees). In that case, the mathematical structure corresponds to an ultrametric space. Ultrametric spaces are ubiquitous in mathematical morphology, ranging from watershed segmentation (minimum spanning tree~\cite{Meyer14,Najman13}) to connected-components preserving filtering (notion of max/min-tree~\cite{Salembier00}).
When an image, or any other element in a dataset embedded into an edge-weighted graph, is represented by a dendrogram, the use of deep learning techniques and in particular of ultrametric convolutional neural networks requires to be have a specific definition of the typical layers for dendrograms: their structure is very different from metric graphs. 
Classical image/data processing opeartors and transforms, including Gaussian and Laplacian operators, convolution, morphological semigroups, etc., have been formulated on ultrametric spaces~\cite{Angulo17,Angulo19}, and the basic ingredients are the ultrametric distance and the distribution of diameters of the ultrametric balls.

\begin{quote}
How to efficiently learn the convolution operation on ultrametric convolutional neural networks? How to define the max-pooling and unpooling-layers on ultrametric convolutional neural networks? Are there specific layers useful in this kind of neural networks?
\end{quote}

Note that this is different from the problem of learning an ultrametric distance from a dissimilarity graph, an optimization problem that can be now efficiently solved by gradient descent methods~\cite{Chierchia20}, useful in learning watershed image segmentation or in other hierarchical clustering methods. Both approaches could be potentially integrated into end-to-end learnable frameworks, where graphs can be embedded into dendrograms and then, ultrametric deep learning techniques could be used for classification or prediction.

%%%%%%%%%%%%%%%%%%%%%%%%%%%%%%%%%%%%%%%%%%%%%%%%%%%%%%%%%%%%%%%%%
\subsection{The difficulty of training morphological neural networks}

The training of morphological neural networks has been, and still is, the object of active research. The dendrite morphological neurons have been trained using geometrical-based algorithms (enclosing patterns in hyper-
boxes)~\cite{Ritter03,Sossa14}. The non-differentiability of the lattice-based operations makes training morphological neural networks more difficult for gradient-based algorithms, but the back-propagation can be achieved without difficulty~\cite{Zhang19}: in fact, maxpooling and ReLU are just examples of morphological layers widely used in deep learning.

More recently, inspired from the tropical mathematics framework, training algorithms not rooted in stochastic optimization but rather on Difference-of-Convex Programming have been used for training dilation and erosion layers~\cite{Charisopoulos17}. These techniques have been also adapted to train other generalized dilation-erosion perceptrons combined with linear transformations~\cite{Valle20}. Additional progress have been made the use of tropical geometry tools for neural network pruning~\cite{Smyrnis19}. The property of sparsity induced by max-plus layers~\cite{Zhang19,Smyrnis19} seems of great potential, however to integrate morphological layers into complex deep learning architectures is not always straightforward from the viewpoint of the optimization.

\begin{quote}
Is there any efficient network learning technique which can combine the optimization techniques inspired from tropical geometry and stochastic gradient descent? 
\end{quote}

\begin{quote}
What are the best gradient-descent optimizers in the case of hybrid networks including morphological layers? Is there any strategy of alternate optimization between the convolution layers and the morphological layers?
\end{quote}

An alternative is to use smooth approximations of the max-min based morphological operators, either using counter-harmonic means~\cite{Masci13} or Log-Sum-Exp terms~\cite{Calafiore18} (related to the Maslov dequantization~\cite{Litvinov07,Angulo13}). In these differentiable frameworks, the morphological ones are the limit cases and it allows, if the nonlinearity parameter is learned too, to provide layers which after training can behave as standard convolutions or as morphological ones.

\begin{quote}
Which smooth approximation to morphological operators is more relevant for training deep learning hybrid networks?
\end{quote}

Discrete geometry focusses on the mathematically sound definitions and efficient algorithms to study binary discrete objects, like lines, circles, convex shapes, etc., as well as to work on discrete functions. Discrete convolution and its equations have been also studied~\cite{Kiselman15}. Deep Learning with limited numerical precision~\cite{Gupta15} and in particular with integer-arithmetic-only~\cite{Jacob17} is relevant in the field of resource-efficient AI, especially for its deployment in embedded systems. The issue of neural network training with constrained integer weights was considered also in the past~\cite{Plagianakos99}. The limit case of discrete representation and computation corresponds to the binary or ternary neural networks~\cite{Courbariaux16,Alemdar17,Zhu17}. Efficiently training these systems uses rounding off, and other numerical tricks, which do not always guarantees some properties related to discrete or binary convolution, down/up-sampling, etc. 

\begin{quote}
Is it possible to incorporate genuine discrete layers and to train them using gradient-descent approaches? Are there other approaches from discrete optimization better adapted to these networks?
\end{quote}

Connections between convolution, non-linear operators and PDEs have been the object of major research in image processing during many years. Image data is interpreted as the discretization of multivariate functions and the output of image processing algorithms as solutions to some PDEs. 
A few works have considered network layers as a set of PDE solvers, where the geometrically significant coefficients of the equation become the trainable weights of the layer~\cite{Ruthotto18,Shen20}. The well-established theory of PDEs allows the introduction of neural network layers with good approximation properties (relevant for the problem of equivariance, for instance). The potential interest of the interpretation of some non-linear layers from the viewpoint of morphological PDEs has been considered in~\cite{Smets20}, even if the formulated PDEs are solved using the viscosity solutions, which correspond to the morphological convolutions. It would be interesting to explore the interest of the numerical solvers for Hamilton--Jacobi PDEs which can be plugged into a deep learning pipelines to learn non-linearities and morphological layers. 

\begin{quote}
What numerical schemes for morphological PDEs are relevant in order to learn morphological operators or another non-linear layer? How to deal with the iterative nature of the approximations?
\end{quote}

I think about the case of deep learning for positive definite matrices (SPD)~\cite{Huang16}. Riemannian geometry-based tools are used to formulate neural network layers which allow computing and learning in that setting. The non-linearity layers, like ReLU and max-pooling, are not well formulated in the SPD case. There is a well-established theory of numerical solution schemes of PDE-based morphology for matrix fields~\cite{Burgeth09,Burgeth17}.

\begin{quote}
Can the numerical schemes for morphological PDEs of matrix fields be used to learn non-linearities for SPD, or other matrices, in Riemannian deep learning? 
\end{quote}

%%%%%%%%%%%%%%%%%%%%%%%%%%%%%%%%%%%%%%%%%%%%%%%%%%%%%%%%%%%%%%%%%
\subsection{Morphological AI}

More than thirty year ago, Schmitt~\cite{Schmitt89} showed the possibility of creating an automatic programming system of artificial intelligence morphology for image processing, in a rule-based paradigm of geometric reasoning. The starting point is the obvious fact that complex morphological transforms are just based on a few bricks. Those primitives can be seen as the words of morphological language and the possible combinations making sense as the grammar of the language. Typical primitives in~\cite{Schmitt89}  are dilations, erosions, hit-or-mis transforms, thinnings and thickennigs, with their corresponding structuring elements. The appropiate constructive combinations provide the grammar. The solution for a given problem is solved using combinational optimization. Other paradigms producing successful automatic design of morphological operators were based on genetic algorithms~\cite{Harvey96} or on PAC (Probably Approximately Correct) learning~\cite{Barrera00}. All these approaches are based on learning the transformations by collections of observed-ideal pairs of images and the result of the desirable operators from these data, which fit with training from datasets as it is done nowadays in deep learning. In the field of deep learning, the first tentative of learning pipelines of morphological operators was~\cite{Masci13}. Using the counter-harmonic mean as an asymptotic approximation to both dilation and erosion, it was proved that stochastic gradient descent-based convolutional neural networks can learn both the structuring element and the composition of operators, including compositions of openings and closings which approximate TV-regularization. However, this kind of approach does not exploit the vast complexity of morphological language. 

An alternative would be to use natural language processing (NLP) deep learning techniques. In that context, the training dataset will be composed of examples of morphological programs, considered as a morphological text, written by experts to solve specific tasks, illustrated with input-output images too. Supervised NLP tasks are based on building pretrained representations of the distribution of words (called word embeddings), such word2vec or context2vect~\cite{Mikolov13,Melamud16}.

\begin{quote}
What is the most appropiate coding of the morphological language to be use with NLP techniques? Is it efficient to use a word for the operator and a declination for the structuring element? Or to use different words for operator and structuring element?
\end{quote}

\begin{quote}
What is the optimal granularity on the decomposition of the operators to obtain a language with a high enough semantic interpretation and which can be still learnt?
\end{quote}

In general NLP, training the word embeddings required a (relatively) large amount of data, which reduced the amount of labeled data necessary for training on the supervised tasks.

\begin{quote}
Do we have a large enough corpus of morphological programs available to learn the operator2vect embeddings? How to extract and parse morphological programs available in multiple repositories to train the algorithms? 
\end{quote}

Another source of inspiration for developing morphological AI is the field of deep coding~\cite{Balog17}. This time, the perspective from the computational morphology viewpoint is to start from a toolbox of programmed morphological functions and the goal would be to learn to write programs using that basic functions. Solving automatic programming problems from input-output examples using deep learning is quite challenge and works only on domain specific languages, which is the case of a morphological language toolbox. It requires to find consistent programs by searching over a suitable set of possible ones, and a ranking of  them if there are multiple programs consistent with the input-output examples. The current paradigm provides only limited and short programs. These techniques can be more efficiently use eventually to rewrite programs and to decompose them into subprograms which can be learn and optimize separately. 

\begin{quote}
How can one introduce syntactical transformations to simplify and rewrite a morphological program?
\end{quote}

\begin{quote}
How to integrate combinatorial and deep learning techniques to explore large enough program spaces which can provide an efficient morphological AI?
\end{quote}

%%%%%%%%%%%%%%%%%%%%%%%%%%%%%%%%%%%%%%%%%%%%%%%%%%%%%%%%%%%%%%%%%%
%\subsection{An image and its multimodal morphological representations}

\section{My conclusion: invest yourself in studying the theory}

I anticipate that the main reaction of the reader may be that of frustration. I have not provided many more reasons for the choice of the previous topics than my own intuition.
I concede that the answer to my open questions will not always lead to any methodological or algorithmic breakthroughs. However, working on the program I discussed above, or on  alternative problems, will definitely advance our discipline and keep it flourishing. 

I am open to collaborate and discuss with anyone interested in some of these topics. Joining efforts to work on challenging problems is fundamental in a research field quickly moving forward and in multiple directions.

A last take-home message. Successful interaction between morphology and deep learning is not only related to computational based aspects. It requires, about all, an intimate understating of the theoretical aspects of both fields. My main advice for young researchers starting a PhD thesis on exploring these interactions is the following: invest part of your time in studying the theoretical papers, read your \emph{classics}, enlarge the scope of your theoretical interests. It will be worthwhile.

%======================================================================
\bibliographystyle{splncs}
%\bibliography{copybibthesis}

\end{document}